\documentclass[aps,prb,twocolumn,superscriptaddress,showpacs]{revtex4}
\usepackage{amssymb}
\usepackage{amsfonts}
\usepackage{amsmath}
\usepackage{graphicx}
\begin{document}


\title{Spin correlations and colossal magnetoresistance in HgCr$_2$Se$_4$}

\author{Chaojing Lin}
\author{Changjiang Yi}
\author{Youguo Shi}
\email{yghsi@iphy.ac.cn}
\affiliation{Beijing National Laboratory for Condensed Matter Physics, Institute of Physics, Chinese Academy of Sciences, Beijing 100190, China}

\author{Lei Zhang}
\affiliation{High Magnetic Field Laboratory, Chinese Academy of Sciences, Hefei 230031, China}

\author{Guangming Zhang}
\affiliation{Department of Physics, Tsinghua University, Beijing 100084, China}

\author{Jens M\"uller}
\affiliation{Institute of Physics, Goethe-University Frankfurt, 60438 Frankfurt (Main), Germany}

\author{Yongqing~Li}
\email{yqli@iphy.ac.cn}
\affiliation{Beijing National Laboratory for Condensed Matter Physics, Institute of Physics, Chinese Academy of Sciences, Beijing 100190, China}
\affiliation{School of Physical Sciences, University of Chinese Academy of Sciences, Beijing 100190, China}
\affiliation{Beijing Key Laboratory for Nanomaterials and Nanodevices, Beijing 100190, China}

\begin{abstract}

This study aims to unravel the mechanism of colossal magnetoresistance (CMR) observed in \emph{n}-type HgCr$_2$Se$_4$, in which low-density conduction electrons are exchange-coupled to a three-dimensional Heisenberg ferromagnet with a Curie temperature $T_C\approx$ 105\,K. Near room temperature the electron transport exhibits an ordinary semiconducting behavior. As temperature drops below $T^*\simeq$ 2.1\,$T_C$, the magnetic susceptibility deviates from the Curie-Weiss law, and concomitantly the transport enters an intermediate regime exhibiting a pronounced CMR effect before a transition to metallic conduction occurs at $T<T_C$. Our results suggest an important role of spin correlations not only near the critical point, but also for a wide range of temperatures ($T_C<T<T^*$) in the paramagnetic phase. In this intermediate temperature regime the transport  undergoes a percolation type of transition from isolated magnetic polarons to a continuous network when temperature is lowered or magnetic field becomes stronger.

\end{abstract}


\maketitle

\section{Introduction}
Colossal magnetoresistance (CMR), the negative magnetoresistance (MR) with magnitude much larger than those observed in conventional ferromagnetic metals, is among the most studied magnetotransport phenomena in condensed matter.~\cite{Coey99,Dagotto01,Salamon01,Tokura02} The MR ratio, defined as $\rho_\mathrm{xx,0}/\rho_\mathrm{xx}(H)$, serves as a figure of merit for CMR materials, where $\rho_\mathrm{xx}(H)$ is the longitudinal resistivity in applied magnetic field $H$ and $\rho_\mathrm{xx,0}$ is the zero-field value. In the past five decades, massive research effort has produced many types of CMR materials, including perovskite manganites,~\cite{Coey99,Dagotto01,Salamon01,Tokura02,Helmolt93,SJin94} europium chalcogenides, monoxide, and hexaboride,~\cite{Molnar67,Methfessel68,Molnar07,Penney72,Shapira73,Sullow00,Das12} chromium spinels,~\cite{Lehmann67} pryochlores,~\cite{Ramirez97,Majumdar98} and cobaltites.~\cite{JWu05} MR ratios up to several orders of magnitude have been obtained.

The CMR-related research has greatly advanced the knowledge of magnetism, electron correlations, and phase transitions~\cite{Coey99,Dagotto01,Salamon01,Tokura02}. Important concepts, such as magnetic polarons,~\cite{Methfessel68,Kasuya70a,Mauger86,Molnar07} and magnetic phase separation,~\cite{Gorkov98,Moreo99} have been developed. However, the physics underlying the CMR effects is often very complicated because of various material-specific complications and many-body interactions. Such complexity is probably best manifested in mixed-valence perovskite manganites, in which many degrees of freedom, including spin, charge, orbital and lattice, come into play, along with other factors, for instance substitutional disorder and strong electron correlations.~\cite{Coey99,Dagotto01,Salamon01,Tokura02} In order to gain further insight into the CMR effects, it is desirable to study a simpler system in which the various experimental parameters are considerably less intertwined.

In this article we demonstrate that \emph{n}-type HgCr$_2$Se$_4$ is an excellent system to study the CMR effect. The ferromagnetism in this material arises from the superexchange interaction between Cr$^{3+}$ ions, which have three 3\emph{d} electrons and thus are free from the Jahn-Teller distortion. The magnetic properties of the Cr$^{3+}$ lattice are barely influenced by conduction electrons, and the critical exponents are found very close to those of an ideal 3D Heisenberg ferromagnet. The paramagnetic-ferromagnetic phase transition drives a well-defined insulator-metal transition with MR ratios up to five orders of magnitude. Most importantly, the CMR effect is most pronounced when spin correlations between the Cr$^{3+}$ ions are significant. Based on these observations, we suggest that spin correlations play an important role in the transport processes related to the magnetic polarons, which are responsible for the CMR effect in a broad range of temperatures in the paramagnetic phase.

HgCr$_2$Se$_4$ has been known as a ferromagnetic semiconductor of the spinel family for several decades.~\cite{Baltzer66,Arai73,Goldstein78,Selmi87,Solin08}
Ferromagnetic order in HgCr$_2$Se$_4$ originates from the Cr$^{3+}$-Se$^{2-}$-Cr$^{3+}$ superexchange interactions between nearest-neighbors of Cr ions.~\cite{Baltzer66}
For \emph{n}-type HgCr$_2$Se$_4$ single crystals with electron densities on the order of 10$^{18}$\,cm$^{-3}$, our recent Andreev reflection spectroscopy experiment has provided evidence for nearly full electron spin polarization in the ferromagnetic ground state.~\cite{GuanT15} In the present work, we focus on the magnetism near the ferromagnetic-paramagnetic phase transition, the metal-insulator transition and the microscopic origin of the CMR effect in \emph{n}-HgCr$_2$Se$_4$.

\section{Experimental methods}
Single crystals of HgCr$_2$Se$_4$ were grown with a chemical vapor method by using CrCl$_3$ or AlCl$_3$ as the carrier agent~\cite{Arai73,Wang13}. The starting materials Hg, Cr, and Se, as well as the carrier agent, were sealed into a silica tube. They were heated at 800\,$^\circ$C for 5-10 days in a tube furnace. A temperature gradient of about 100\,$^\circ$C was used for the transport at the cooler growth end of the tube. The obtained octahedral-shaped crystals typically have sizes in a range of 0.5 to 2\,mm. Electron transport measurements were carried out in helium vapor flow cryostats with a temperature range from 1.5 to 325\,K and magnetic fields up to $\mu_0H$=14\,T. The magnetization of HgCr$_2$Se$_4$ single crystals were measured in a commercial vibrating sample magnetometer with a temperature range of 1.8-400\,K and magnetic fields up to 7\,T.  Details of transport and magnetization measurements, as well as the method for extracting critical exponents are given in Appendix\,A.

\section{Experimental results}

Our measurements have been carried out on several batches of \emph{n}-type HgCr$_2$Se$_4$ single crystals. All of the data shown in the main text were taken from a sample with an electron density of $\sim1\times10^{18}$\,cm$^{-3}$ at liquid helium temperatures. We have also obtained similar results from other HgCr$_2$Se$_4$ samples with comparable carrier densities. Additional magnetization and electron transport data are given in Appendices\,B-D.

\begin{figure}
\includegraphics[width=8.5 cm]{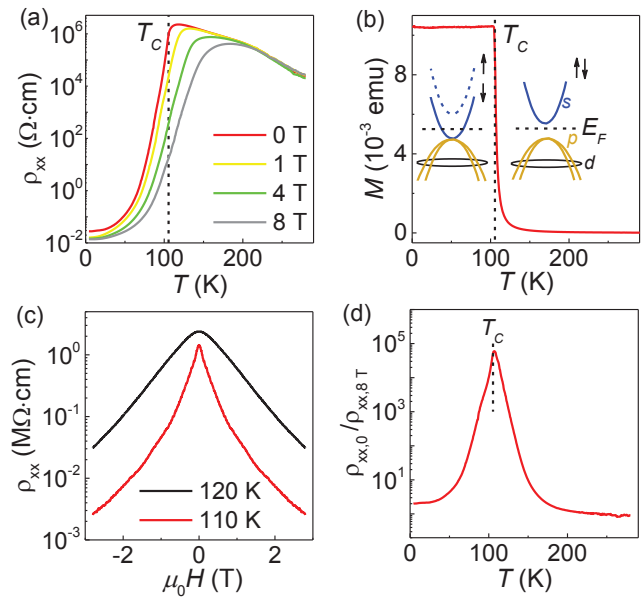}
\caption{(color online). (a) Temperature dependences of longitudinal resistivity $\rho_\mathrm{xx}$ for an \emph{n}-type HgCr$_{2}$Se$_{4}$ sample with carrier density $n\approx1\times10^{18}$\,cm$^{-3}$ for magnetic fields $\mu_0 H$= 0, 1, 4 and 8\,T. The corresponding Hall measurement results are shown in Fig.\,\ref{Fig_Hall} in Appendix\,A1. (b) $T$-dependence of the magnetization $M$ with an applied field $\mu_0 H$=0.01\,T. The left (right) inset is a sketch of the position of the spin-split (spin-degenerate) conduction band for temperatures below (above) the Curie temperature. (c) $\rho_\mathrm{xx}$ plotted as a function of $\mu_0 H$ for $T$=110\,K and 120\,K. (d) Temperature dependence of the MR ratio $\rho_\mathrm{xx,0}/\rho_\mathrm{xx}(H)$ with $\mu_0 H$=8\,T.The maximum is located at $T$=106.9\,K. The Curie temperature is marked with a vertical dashed line in panels (a), (b) and (d).}
\end{figure}

Fig.\,1(a) shows the temperature dependences of the longitudinal resistivity $\rho_\mathrm{xx}$ in magnetic fields $\mu_0 H$=0, 1, 4 and 8\,T. At $T$=4.2\,K the electron mobility is about $4\times10^2$\,cm$^2$V$^{-1}$s$^{-1}$. For temperatures below $\sim$\,60\,K, $\rho_\mathrm{xx}$ increases slowly with increasing $T$. In contrast, a very rapid increase of $\rho_\mathrm{xx}$ is observed as the temperature is further raised. At $T=118$\,K, $\rho_\mathrm{xx}$ reaches a maximum, with a value about eight orders of magnitude larger than the low temperature values. Such a large temperature dependence is several orders magnitude larger than those observed by other groups in HgCr$_{2}$Se$_{4}$.~\cite{Goldstein78,Selmi87,Solin08} This clearly demonstrates a metal-insulator transition driven by the ferromagnetic-paramagnetic phase transition shown in Fig.\,1(b).~\cite{Coey99,Penney72,GuanT15} The semiconducting behavior in the paramagnetic phase can be attributed to the Fermi level being located below the conduction band minimum, whereas in the ferromagnetic phase a large spin splitting (nearly 1\,eV) shifts the Fermi level above the bottom of the spin-down conduction band because of strong exchange interactions between Hg-6s and Cr-3d electrons [See insets in Fig.\,1(b)].~\cite{Arai73,GXu11}  At temperatures near the Curie temperature ($T_C\approx\,$105\,K), especially on the paramagnetic side, giant negative magnetoresistances are observed, as illustrated in Fig.\,1(c-d). The temperature dependence of the MR ratio is very sharp, with a maximum of $6\times10^4$ at $T$=106.9\,K and $\mu_0 H$=8\,T. Such a large MR ratio is comparable to the highest values reported in literature (e.g. $\sim10^6$ in europium monooxide,~\cite{Penney72} and $\sim10^3$-10$^5$ in manganites,~\cite{SJin94,Chen96}) and is at least two orders of magnitude larger than previously reported values for HgCr$_2$Se$_4$.~\cite{Goldstein78,Solin08} Another noticeable feature in Fig.\,1 is that the maxima of both $\rho_\mathrm{xx}$ and the MR ratio are located at temperatures above $T_C$, suggesting that the metal-insulator transition and the magnetic phase transition are not fully synchronized.

\begin{figure}[t]
\includegraphics[width=8.5 cm]{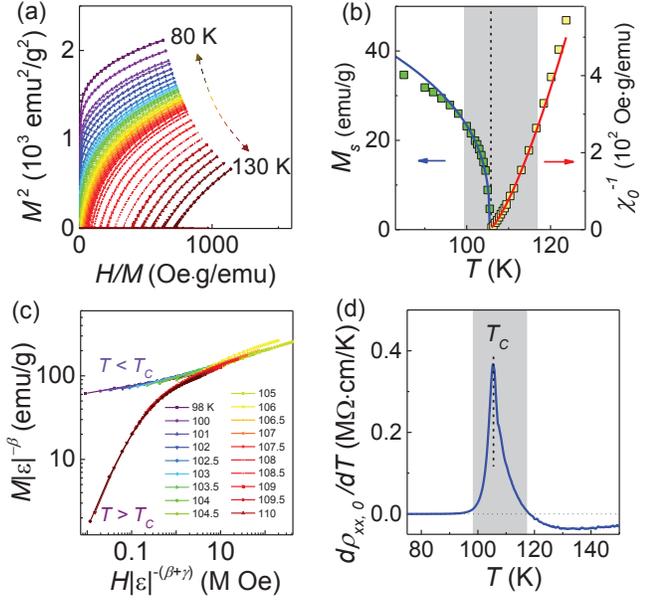}
\caption{(color online). (a) Arrott plot of the magnetization isotherms ($M^2$\,vs.\,$H/M$) in a temperature region of 80-130\,K. The temperature interval is 0.5\,K for 100-110\,K and 2\,K for the others. (b) $T$-dependences of the spontaneous magnetization $M_s$ (green squares, left) and the zero-field limit of the inverse magnetic susceptibility 1/$\chi_0$ (yellow squares, right). Also shown are fitting curves (solid lines) to the critical equations. (c) Scaling plot of $M|\varepsilon|^{-\beta}$\,vs.\,$H|\varepsilon|^{-(\beta+\gamma)}$ for $T$=98-110\,K, where $\varepsilon=(T-T_C)/T_C$ is the reduced temperature. (d) $T$-dependence of the zero-field resistivity derivative $d\rho_{\mathrm{xx,\,0}}/dT$. In panels (b) and (d), the critical region is indicated with a gray zone.}
\end{figure}

More insight into the magnetism in HgCr$_2$Se$_4$ can be gained from an analysis of the critical exponents.~\footnote{The method for the analysis of critical exponents is given in Appendix\,A2.} The Arrott plot in Fig.\,2(a) allows for an extraction of the critical exponents from the magnetization data near $T_C.$~\cite{Fisher67,Arrott67} The critical exponents $\beta$, $\gamma$ and $\delta$ can be obtained from fits to the following asymptotic relations: $M_s(T)=\mathrm{lim}_{H\rightarrow0}M(T)\propto[(T_C-T)/T_C]^\beta$ at $T<T_C$, $\chi_0(T)=\mathrm{lim}_{H\rightarrow0}M(T)/H\propto[T_C/(T-T_C)]^\gamma$ at $T>T_C$ and $M\propto H^\delta$ at $T=T_C$. Here $M_s$ is the spontaneous magnetization, and $\chi_0$ is the zero-field limit of magnetic susceptibility. As shown in Fig.\,2(b) and Appendix\,A2, these fits yield $\beta$=0.361, $\gamma$=1.372, and $\delta$=4.84, which deviate strongly from the mean-field values ($\beta$=1/2, $\gamma$=1, and $\delta$=3), but are close to the theoretical values for a 3D Heisenberg ferromagnet ($\beta$=0.367, $\gamma$=1.388, and $\delta$=4.78).~\cite{Blundell01} The validity of the critical exponents analysis can be verified by the scaling plot shown in Fig.\,2(c). All of the data in the critical region collapse very nicely onto two curves in the $M/|(T-T_C)/T_C|^\beta$\,vs.\,$H/|(T-T_C)/T_C|^{\beta+\gamma}$ plot: one for $T<T_C$  and the other for $T>T_C$. Such a scaling behavior is expected for the critical region of a second-order magnetic phase transition.~\cite{Fisher67} It is important to note that the magnetic properties are dominated by the superexchange-coupled Cr$^{3+}$ ions,~\cite{Arai73} since the conduction electrons are outnumbered by Cr$^{3+}$ ions by four orders of magnitude.

From the fits for extracting critical exponents $\beta$ and $\gamma$, we obtain $T_C$=105.35\,K, which is close to the maximum in the temperature derivative of the zero-field resistivity, $d\rho_{\mathrm{xx},\,0}/dT$, as shown in Fig.\,2(d). The sharp peak in $d\rho_{\mathrm{xx},\,0}/dT$ coincides very well with the critical region (98-117\,K), determined from the $M_s(T)$\,vs.\,$T$ and $\chi_0^{-1}(T)$\,vs.\,$T$ plots in Fig.\,2(b). This implies that, due to the \emph{s-d} exchange interaction, the critical behavior of the magnetic lattice is corresponding to the strongest change in the electron transport properties. Sharp peaks in $d\rho_{\mathrm{xx},\,0}/dT$ at $T=T_C$ have also been observed in ferromagnetic metals.~\cite{Craig67,Yanagihara02} The critical behavior in case of metallic conductivity was explained long ago by Fisher and Langer,~\cite{Fisher68} who pointed out that short-range spin fluctuations are responsible for the singular behavior in $d\rho_{\mathrm{xx},\,0}/dT$ at $T=T_C$. The Fisher-Langer model, however, cannot be used to describe the peak in $d\rho_{\mathrm{xx},\,0}/dT$ observed in this work, since $\rho_\mathrm{xx}$ in \emph{n}-HgCr$_2$Se$_4$ is on the order of 10$^5\,\Omega\cdot$cm or higher near $T_C$, suggesting the transport is far from the metallic regime.~\footnote{For $n\sim10^{18}$\,cm$^{-3}$, $\rho_\mathrm{xx}$ less than $\sim0.1$\,$\Omega\cdot$cm is required to meet the Ioffe-Regel criterion for metallic conduction.}

\begin{figure}
\includegraphics[width=8.5 cm]{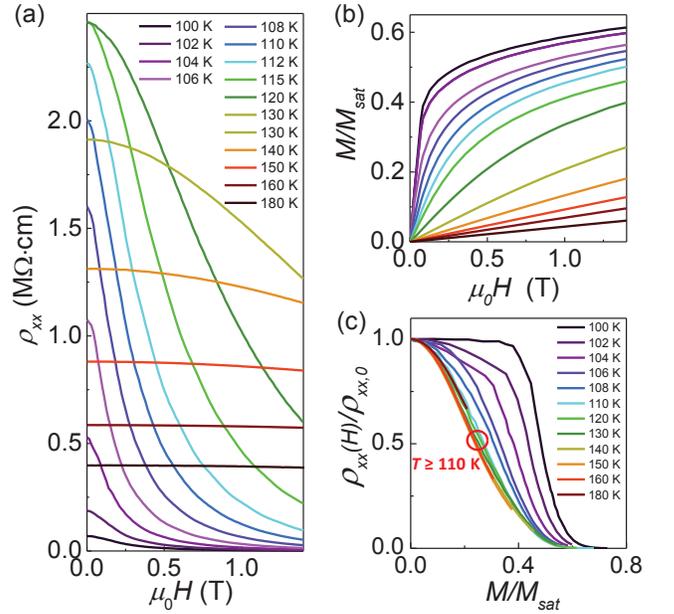}
\caption{(color online). (a) Magnetic field dependence of longitudinal resistivity $\rho_{\mathrm{xx}}$, (b) Dimensionless magnetization $m(H)=M(H)/M_\mathrm{sat}$, where $M_\mathrm{sat}$ is the saturation magnetization (equivalent to 3\,$\mu_B/$Cr$^{3+}$). (c) Normalized resistivity, $\rho_{\mathrm{xx}}(H)/\rho_{\mathrm{xx,\,0}}$ plotted as a function of $m(H)$. The data for $T\geq$110\,K can be roughly scaled onto a single curve. The same set of temperatures (100-180\,K) are used in panels (a) and (b)}
\end{figure}

Fig.\,3(a) shows a set of $\rho_\mathrm{xx}(H)$\,vs.\,$H$ curves for $T$=100-180\,K, at which the CMR effect is most pronounced. The corresponding magnetization curves are shown in Fig.\,3(b). In order to reveal the connection between transport and the magnetization, in Fig.\,3(c) we plot the normalized resistivity, defined as $\rho_\mathrm{xx}(H)$/$\rho_\mathrm{xx,\,0}$, as a function of reduced magnetization, $m(H)$=$M(H)/M_{sat}$, where $M_\mathrm{sat}$ is the saturation magnetization corresponding to 3\,$\mu_B/$Cr$^{3+}$. It is found that the curves for $T\geq110$\,K can be approximately scaled onto a single curve. In contrast, no such scaling behavior exists for $T<110$\,K.

The scaling behavior between the MR and the magnetization implies that a unified description of the electron transport is possible for a broad range of temperatures in the paramagnetic phase. It is connected with a transport regime in an intermediate temperature zone illustrated in Fig.\,4. At higher temperatures [$T>T^\ast\approx2.1T_C\approx\,$220\,K, zone $\mathrm{I}$ in Fig.\,4(a)], the transport follows a thermal activation law $\rho_\mathrm{xx}\propto \mathrm{exp}(\Delta/k_BT)$ with $\Delta$=0.19\,eV, whereas at lower temperatures ($T<T_C$, zone $\mathrm{III}$), $\rho_\mathrm{xx}$ drops very rapidly with decreasing $T$ and a transition to metallic conduction takes place. In the intermediate temperature zone ($T_C<T<T^\ast$, zone $\mathrm{II}$), the resistivity exhibits a crossover behavior between the high and low temperature regimes. These three transport regimes are \emph{correlated} remarkably well with the behavior of the $T$-dependence of the inverse magnetic susceptibility $\chi^{-1}(T)$ plotted in Fig.\,4(b). In zone $\mathrm{I}$, $\chi^{-1}(T)$ follows a Curie-Weiss law, and $\rho_\mathrm{xx}$ is nearly independent of the magnetic field [Fig.\,4(a)]. The sample behaves like an ordinary semiconductor with activated transport. As temperature is lowered below $T^\ast$ (zone $\mathrm{II}$), the susceptibility deviates from the Curie-Weiss law.~\footnote{Determination of $T^\ast$ from the susceptibility data is shown in Fig.\,\protect{\ref{Fig_susceptibility}}.} The zero-field resistivities in zone $\mathrm{II}$ are much smaller than extrapolated values from the thermal activation law extracted in zone $\mathrm{I}$. When a magnetic field is applied, $\rho_\mathrm{xx}$ in zone $\mathrm{II}$ drops more strongly at the lower temperature side.
As a consequence, the intermediate transport regime is squeezed to a narrower range of temperatures in stronger fields, as shown in Fig.\,4(a).

\begin{figure}
\includegraphics[width=8.5 cm]{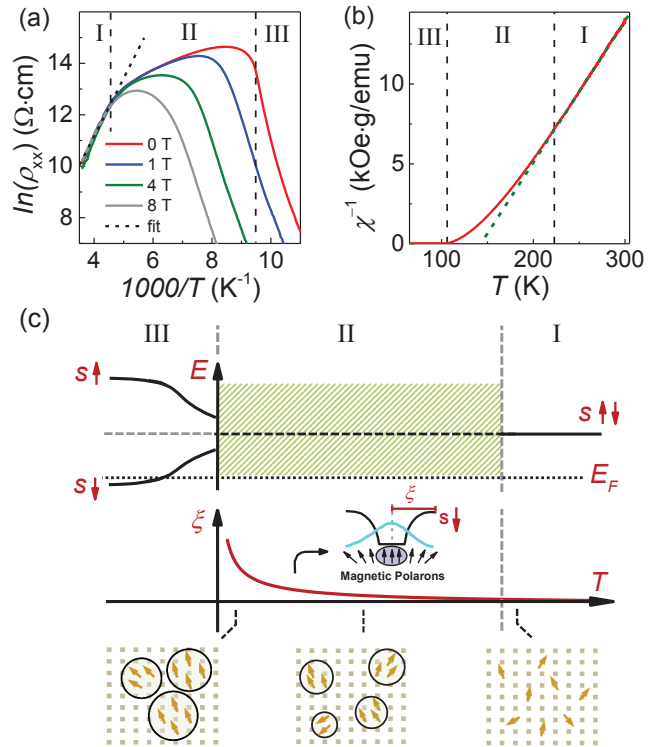}
\caption{(color online). (a) Temperature dependence of the resistivity in the form of ln($\rho_\mathrm{xx}$)\,vs.\,1000/$T$ for magnetic fields $\mu_0 H$=0, 1, 4, 8\,T. The linear fit (dotted line) for $T>\,$220\,K yields a thermal activation gap $\Delta$=0.19\,eV. (b) $T$-dependence of the inverse susceptibility $\chi^{-1}$ recorded at $\mu_0 H$=0.01\,T (solid line), which starts to deviate from the Curie-Weiss law (dotted line) at $T^\ast\approx$\,220\,K (See Appendix\,B1 for details). (c) The upper panel schematically shows the evolution of the \emph{s}-orbital states with decreasing temperature. In zone $\mathrm{I}$ ($T>T^\ast$) the conduction band is spin-degenerate, whereas in zone $\mathrm{III}$ ($T_C<T<T^\ast$) the ferromagnetic order causes a large spin-splitting of the conduction band. The middle panel shows that spin correlation length $\xi$ becomes longer with decreasing temperature. In zone II, $\xi$ enters the nanometer scale and can substantially increase the size of magnetic polarons, as illustrated in the bottom panel.}


\end{figure}

\section{Discussion}

A crucial point of our experimental observation is the existence of the intermediate transport regime and its correspondence with the deviation of magnetic susceptibility from the Curie-Weiss law in zone II. Such deviation has been well understood and can be attributed to the spin correlations and fluctuations that are neglected in mean-field theories.~\cite{Blundell01,Stanley71} The critical behavior of the magnetic phase transition is a result of a diverging spin correlation length at $T=T_C$. The sharp peak in $d\rho_{\mathrm{xx},\,0}/dT$ near $T_C$ [see Fig.\,2(d)] can be viewed as a manifestation of such spin correlations in electron transport due to the strong \emph{s-d} exchange interaction. In contrast, when the temperature is high enough so that the spin correlations are negligible, the transport exhibits little response to the magnetic field, as illustrated in zone I, Fig.\,4(a). Taking these observations altogether, it is reasonable that the spin correlation effects are important to account for the transport properties observed in zone $\mathrm{II}$.

In this intermediate temperature zone, spin correlation length becomes long enough to favor nanoscale local ordering of Cr$^{3+}$ spins. Neutron scattering measurements have confirmed such spin correlations in the Heisenberg-type ferromagnets (e.g.\ EuO and EuS) at temperatures below $\sim2T_C$.~\cite{Als-Nielsen76} However, the transport properties of these classic magnetic semiconductors were explained in terms of magnetic polaron models \emph{without} considering the spin correlation effects.~\cite{Kasuya70a,Mauger86,Emin87,Majumdar98} The magnetic polaron is a nanoscale charge-spin composite, in which a charge carrier polarizes a number of localized spins in the lattice of magnetic ions via the exchange interaction. The existence of magnetic polarons have been confirmed by a lot of experiments, including muon spin resonance,~\cite{Storchak09,Storchak10} and Raman scattering.~\cite{Snow01,Rho02}  Thermally activated transport has been observed in many magnetic semiconductors with low carrier densities, such as EuO,~\cite{Penney72,Shapira73} CdCr$_2$Se$_4$,~\cite{Prosser74} Sc-doped Tl$_2$Mn$_2$O$_7$.~\cite{Ramirez97} The resistivity follows $\rho_\mathrm{xx}\sim\exp(\Delta_b/k_B T)$, mostly in a temperature range of a few times $T_C$ to room temperature, where $\Delta_b$ is attributed to the binding energy of individual magnetic polarons, typically of order 0.1\,eV. At lower temperatures, the transport was found to deviate from the thermally activated transport with a single energy barrier. Up to now, not much effort has been made to explain the lower resistivities in this temperature range quantitatively, except a work reported by Majumdar and Littlewood.~\cite{Majumdar98}  The equation for $\Delta_b(T)$ given there, however, predicts larger slopes of $\ln\rho_\mathrm{xx}(T)$ at lower temperatures, inconsistent with the data shown in Fig.\,1(a). It is noteworthy that in Ref.\,\cite{Majumdar98}, as well as other theoretical studies, the exchange interaction between the lattice spins are either treated on a \emph{mean field} level or being \emph{neglected}.~\cite{Kasuya70a,Mauger86,Majumdar98,Molnar07} Spin correlation effects are therefore not included in existing models of magnetic polarons for magnetic semiconductors, to the best of our knowledge. In the following, we suggest that spin correlations account for the transport characteristics observed in zone II.

In the \emph{n}-type HgCr$_2$Se$_4$ samples studied in this work, the charge donors are presumably Se vacancies.~\footnote{This is based on the fact that the samples tend to be $n$-type ($p$-type) after being annealed in Hg (Se) vapor.} Similar to other magnetic semiconductors with anion deficiencies,~\cite{Penney72,Shapira73} electron clouds surrounding unionized donors lead to formation of \emph{bound} magnetic polarons.~\cite{Kasuya70a} The \emph{s-d} exchange interaction (or \emph{s-f} interaction in rare earth compounds) provides strong polarizing force for a small number of lattice spins overlapping with the donor wavefunctions.~\cite{Penney72,Mauger86,Storchak09,Storchak10,Molnar07} At high temperatures, spin correlation length $\xi$ is too short to be relevant. The lattice spins in the polarons can thus be modeled as a uniformly polarized core of sub-nm size, separated by a sharp boundary from the paramagnetic background with randomized spins. This has been the basis of previous theories of bound magnetic polarons.\cite{Kasuya70a,Mauger86,Majumdar98} At lower temperatures, increasing $\xi$ makes this simplified picture no longer valid.  Spin polarization at the core of the magnetic polarons remains large, but it decays over a distance on the order of $\xi$ , rather than abruptly dropping to zero at the boundary.  Fig.\,4(c) illustrates such polarons with the core-shell structure, in which the shell width is determined by the spin correlation length.

The magnetic polarons enlarged by spin correlations can have substantial impact on the electronic structure and transport properties in zone II. As shown in Fig.\,4(c), the \textit{s-d} exchange interaction produces a 3D confinement potential for the spin-down \emph{s}-orbital states. The energy levels in the quantum-dot-like structure are expected to be lower than the spin-degenerate 6\emph{s} band in the paramagnetic phase. This is supported by an optical measurement by Arai et al.,~\cite{Arai73} who observed a 0.3\,eV red shift of the absorption edge at temperatures above $T_C$ and the shift becoming noticeable at about 200\,K. This temperature is close to $T^*$ obtained in this work, further revealing the importance of spin correlations. The empty states in the quantum-dot-like structure provide a new avenue for electron transport. The carriers bound by the donors no longer need to be excited to the spin-degenerate conduction band. This can reduce the activation energy substantially, in agreement with the rapid decrease in the slope of the $\ln\rho_\mathrm{xx}$ vs.\ $1/T$ plot in Fig.\,4(a) at $T<T^*$. When the temperature is further lowered deeply into zone II, increasing $\xi$ causes more overlapping in the wavefunctions of neighboring polarons, and the transport will eventually pass a threshold at which the polarons form a percolated network, as depicted in the bottom panel of Fig.\,4(c).

For the HgCr$_2$Se$_4$ sample with electron density of $\sim10^{18}$cm$^{-3}$ at low temperatures, the nearest neighbor distance between the polarons is about $10$\,nm. A spin correlation length of a few nanometer should be able to establish the percolation. Even though no neutron scattering experiment has been reported for HgCr$_2$Se$_4$, the data from other Heisenberg-type ferromagnets imply that $T-T_C\sim10$\,K is probably sufficient to reach the percolation threshold.
Generally speaking, the percolation type of transition exists in many CMR materials. However, it has been difficult to identify a convenient universal hallmark in transport for the onset of percolation. Nevertheless, in the HgCr$_2$Se$_4$ samples studied in this work, the transport related to the polarons evolves from a well-defined thermally activated transport at higher temperatures (zone I). Very high resistivity ($\sim10^5-10^6$\,$\Omega\cdot$cm) is maintained throughout zone II. The crossover from the isolated polarons to percolation is hence not complicated by additional transport channels (e.~g.\ two-band conductivity in EuB$_6$ reported in Ref.\,\cite{ZhangXH08}). Fig.\,1(a) shows that the zero-field resistivity has a maximum at $T=118$\,K, which is reasonably close to the value expected for the onset of percolation. Therefore, the temperature at which the resistivity maximum appears, $T_\mathrm{Rmax}$, may provide a good estimate for the percolation threshold, $T_p$, in \emph{n}-HgCr$_2$Se$_4$.

When the magnetic field is applied, the resistivity in zone II is greatly reduced. Such negative MR can be attributed to sensitive response of magnetic polarons to  the external field. The Zeeman energy favors magnetic ordering, and consequently increases the core size of magnetic polarons, and therefore decreasing their binding energy. Assuming a constant total size (i.~e.\ core + shell) of the polarons at the percolation threshold, a higher $T_p$ is thus expected for a stronger magnetic field. This is consistent with our experimental data.  As shown in Fig.\,1(a), $T_\mathrm{Rmax}$ increases from 118\,K to 184\,K as the magnetic field is increased from 0 to 8\,T.

\begin{figure}
\includegraphics[width=8.5 cm]{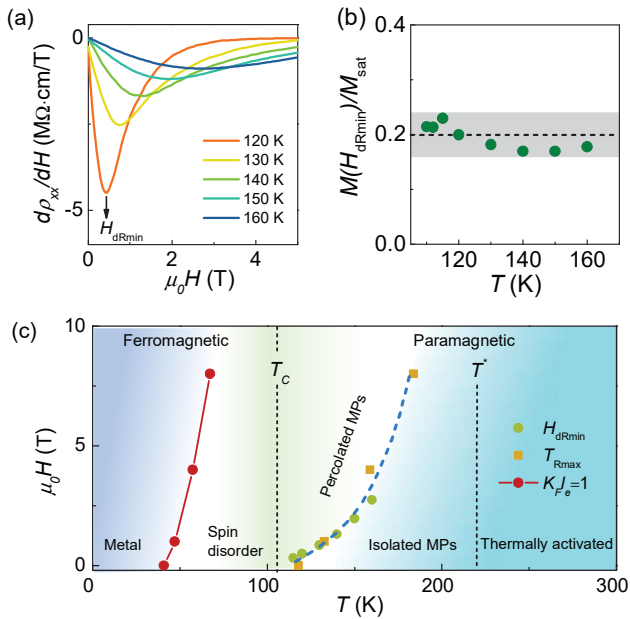}
\caption{(color online). (a) Derivative of resistivity, $d\rho_\mathrm{xx}/dH$, as a function of magnetic field for $T=$120-160\,K. The position of the minimum is denoted as $H_\mathrm{dRmin}$. (b)Magnetization corresponding to the $d\rho_\mathrm{xx}/dH$ minimum, in the units of saturation magnetization, $M_\mathrm{sat}$; (c) Phase diagram in the ($T$, $H$) plane. In zone II, the ($T_\mathrm{Rmax}$,$H$) (squares) and the ($d\rho_\mathrm{xx}/dH$) (circles) data roughly fall onto a single line, which separates zone II into two regions: one with isolated magnetic polarons (MPs), and the other with percolated or merged MPs.
}
\end{figure}

In a recent study of the CMR effect in EuB$_6$, Amyan et al.\ suggested that analysis of the derivative of the MR data provides a convenient method for determining the percolation threshold.~\cite{Amyan13} Based on an experiment in combination of conventional transport, nonlinear transport and noise spectroscopy, they concluded that the minimum in the $d\rho_\mathrm{xx}/dH$ curve can be ascribed to the beginning of percolation. Because of its negative sign, the $d\rho_\mathrm{xx}/dH$ minimum is corresponding to the most sensitive response to the change in magnetic field. Shown in Fig.\,5(a) are a set of $d\rho_\mathrm{xx}/dH$ curves at $T=$120-160\,K for the HgCr$_2$Se$_4$ sample. Each of them has a pronounced dip. The location of the minimum, $H_\mathrm{dRmin}$, increases monotonically with increasing temperature. It is striking that the corresponding magnetization, $M(H_\mathrm{dRmin})$ remains at about 0.2$M_\mathrm{sat}$ for all temperatures, as shown in Fig.\,5(b). Similar analysis of the MR data of EuB$_6$ yielded a critical magnetization of $\sim0.1M_\mathrm{sat}$. It is noteworthy, however, that the spins inside the magnetic polarons (including the shell areas) are partially polarized. This suggests that the actual volume ratio could be considerably larger than 20\% for HgCr$_2$Se$_4$ and 10\% in EuB$_6$ at the percolation threshold. As discussed in previous works on other CMR materials,~\cite{ZhangXH09,Das12,Amyan13,YuLQ13} the transition from the isolated magnetic polarons to the percolated network has some resemblance of granular metal films, in which percolation thresholds of about 0.5 have been observed.~\cite{ZhangXX01} It is however out of the scope of this work to discuss whether the percolation transition is quantum or classical in nature.~\cite{WanCC02}

Fig.\,5(c) shows that the percolation thresholds determined from the two methods described above, namely ($T_\mathrm{Rmax}$, $H$) and ($T$, $H_\mathrm{dRmin}$), agree with each other quite well. This threshold line separates zone II into two electronic phases: one with isolated magnetic polarons at low $H/T$ and the other with percolated or merged polarons at high $H/T$. At high temperatures ($T>T^*$, zone I), the lattice Cr spins are uncorrelated and the transport is thermally activated and is little influenced by the magnetic field. This is in contrast to the regions with the CMR effect, which include nearly the entire zone II and a part of zone I close to $T_C$. It should be noted that the metallic conductivity only appears at temperatures much lower than $T_C$. In Fig.\,5(c), a $k_F l_e\sim1$ line is used as a crude boundary between the metallic and insulating phases, where $k_F$ is the Fermi wavevector, and $l_e$ is the electron's mean free path.~\footnote{The mean free path $l_e$ is estimated with an effective mass of $m^*=0.15m_e$, which was obtained from an optical measurement reported in Ref.\,\protect{\cite{Selmi87}}.} Near $T_C$, the resistivity is very high despite the transport has passed the percolation threshold. This can be attributed to strong spin fluctuations as well as the disorder effect related to the low density electron system. Truly metallic conductivity takes place only when the spontaneous magnetization is large enough such that the bottom of the conduction band drops considerably below the Fermi level to overcome the disorder effect, which arises from defects and spin excitations. Disorder induced localization has been discussed previously by Coey et al.\ in manganites.~\cite{Coey95} It is also likely that the transport involves the hopping of magnetic polarons, which have a very large effective mass due to the strong coupling between the charge carriers and the magnetic lattice. The polaron localization and hopping were discussed long ago by von Moln\'{a}r et al.\ for the high resistivity observed near $T_C$ in Eu-chalcogenides.~\cite{Molnar67,Molnar07}

\section{Conclusion}

In summary, we have shown that spin correlations play a significant role in the transport properties of the CMR material \emph{n}-HgCr$_2$Se$_4$, in which a low-density electron system is exchange-coupled to lattice spins with Heisenberg-type ferromagnetic order. For a wide range of temperatures in the paramagnetic phase, the transport is related to magnetic polarons and their effective size is dependent on the spin correlation length. The colossal negative magnetoresistance (with MR ratio up to five orders of magnitude) can be explained by a percolation type of transition from isolated magnetic polarons to a continuous network, which can be driven either by lowering the temperature or increasing magnetic field. Our work calls for further work to investigate the magnetic polarons in magnetic semiconductors and related materials, in particular with microscopic probes. This will help to gain further insight into the physics of magnetic polarons influenced by spin correlations, which have largely been neglected in previous studies.

\begin{acknowledgements} We are grateful to stimulating discussions with X. Dai, Z. Fang, J. R. Shi, S. von Moln\'{a}r, H. M. Weng, P. Xiong, Y. F. Yang, and L. J. Zou.  This work was supported by the National Basic Research Program of China (Project numbers 2012CB921703 \& 2015CB921102), National Science Foundation of China (Project numbers 11374337, 11474330 \& 61425015), National Key Research and Development Program of China (2016YFA0300600) and the Strategic Initiative Program of Chinese Academy of Sciences (Project numbers XDB070200).
\end{acknowledgements}

\begin{appendix}
\section{ADDITIONAL DETAILS OF EXPERIMENTAL METHODS}

\subsection{Transport measurements}

\begin{figure}[b]
\includegraphics[width=7 cm]{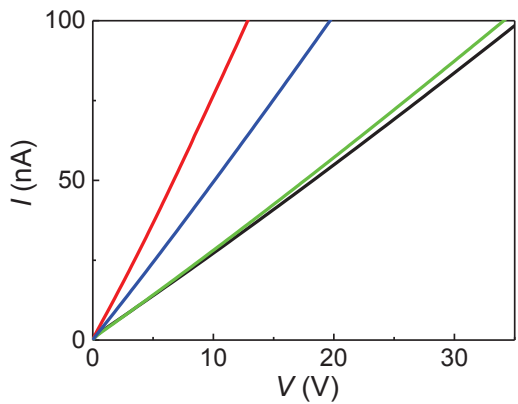}
\caption{\label{Fig_IVcurve}(color online) Current-voltage characteristics at $T$=140\,K for several pairs of electrical contacts on a \emph{n}-HgCr$_2$Se$_4$ sample. The linear $I$-$V$ curves suggest that the contacts are ohmic despite that the impedances are on the order of 100\,M$\Omega$. }
\end{figure}

Electric contacts were made with silver paint or thermally evaporated Cr/Au thin films on a flat (111) surface of HgCr$_2$Se$_4$ single crystals. The samples are mostly 0.5-1\,mm thick. Mechanical thinning or polishing was not used in order to avoid unintentional modifications of transport properties. A multiple-terminal contact geometry was used for the longitudinal resistivity and the Hall resistivity measurements. The distances between the source and drain contacts are larger than or comparable to the sample thicknesses. The standard low frequency lock-in technique was used for the measurements of the low-impedance metallic states, whereas the \emph{dc} method was employed in case of very high impedances (e.g.\ semiconducting states at $T>T_C$). In the \emph{dc} measurements, current pulses of both positive and negative polarities were applied alternatingly so that spurious thermoelectric effects can be removed. The input impedances of preamplifiers and voltmeters were chosen to be much larger than the sample impedances. The magnitude of the applied current was set at a low level in order to keep the transport in the linear regime.  Fig.\,\ref{Fig_IVcurve} shows several examples of the current-voltage ($I$-$V$) curves in the high impedance regime. The $I$-$V$ characteristics was confirmed to be linear for every pair of electrical contacts used in the transport measurements.

\begin{figure}
\includegraphics[width=7 cm]{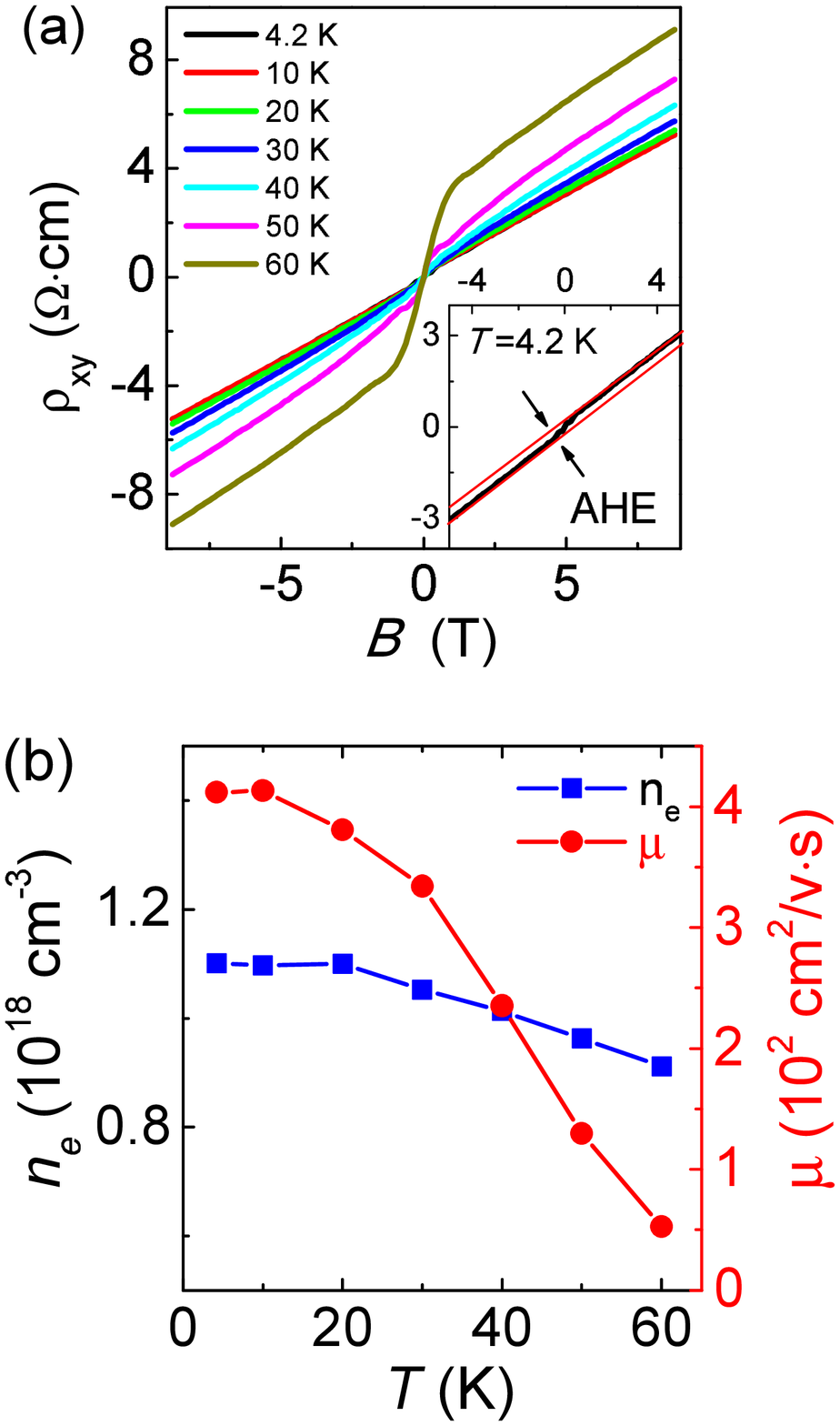}
\caption{\label{Fig_Hall}(color online) (a) The Hall resistivity of Sample A in temperature range from 4.2\,K to 60\,K and in magnetic field up to 9\,T. The inset shows the Hall effect data at $T$=4.2\,K, in which the anomalous Hall effect (AHE) can also be observed. (b) Temperature dependence of electron density $n_e$ and mobility $\mu$.}
\end{figure}

In order to extract the longitudinal and Hall resistances without intermixing, a standard symmetrization/anitisymmetrization was applied the raw transport data with respect to two polarities of the magnetic field. Fig.\,\ref{Fig_Hall}(a) shows the Hall resistivity data of the HgCr$_2$Se$_4$ sample discussed in the main text (Sample A) at selected temperatures from 4.2\,K to 60\,K and in magnetic field up to 9\,T. In low field region, the nonlinear of the Hall resistivity is from the anomalous hall effect which can be attributed to the ferromagnetism of HgCr$_2$Se$_4$. For the curve at $T$=4.2\,K, the anomalous Hall component has been clearly marked at inset of Fig.\,\ref{Fig_Hall}(a) by red lines. While at high magnetic fields (e.g.\ 6-9\,T), the Hall resistivity shows a linear behavior because of the saturation in magnetization. This allows for straightforward extractions of the Hall coefficient $R_H$. The corresponding electron densities, obtained by $n_e=1/eR_H$, are plotted in Fig.\,\ref{Fig_Hall}(b) for several temperatures. The mobility $\mu$ can be obtained by using the relation $\mu=R_H/\rho_\mathrm{xx,\,0}$ where $\rho_\mathrm{xx,\,0}$ is the zero field resistivity [See Fig.\,1(a)].

\subsection{Analysis of critical exponents in magnetization}

\begin{figure}
\centering
\includegraphics[width=7 cm ]{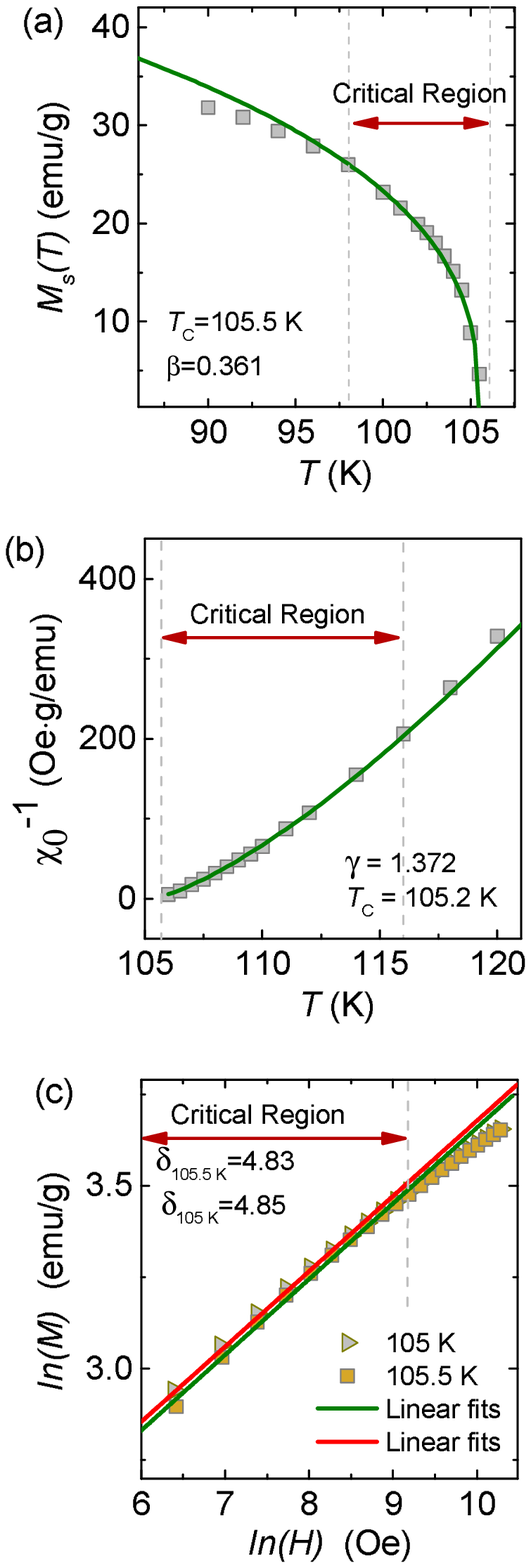}
\caption{\label{Fig_critical_expn}(color online). (a) Spontaneous magnetization $M_s(T)$ (squares) and the fit to the Eq.\,(A3a) (solid line). (b) Zero field inverse magnetic susceptibility 1/$\chi_0$(T) (squares) and the fit to Eq.\,(A3b) (solid line). (c) $M(H)$ at $T$=105\,K (triangles) and 105.5\,K (squares) and the fits to Eq.\,(A3c). Both $M$ and $H$ are in the logarithmic scale.}
\end{figure}

From Landau's mean field theory of phase transitions, the basic thermodynamic potential is Gibbs free energy $G(T,M)$, which is a function of temperature $T$ and magnetization $M$. The latter is generally treated as an order parameter in ferromagnetic systems. Near the paramagnetic to ferromagnetic transition, $G(T,M)$ can be expanded as a power series of $M$:
\begin{eqnarray}
G(T,M)=G_0+\frac{a(T)}{2}M^2+\frac{b(T)}{4}M^4-MH
\end{eqnarray}
where the coefficients $a$ and $b$ are temperature dependent. At equilibrium, $dG/dM$=0, and one obtains:
\begin{eqnarray}
\label{eq_A2}
\frac{H}{M}=a+bM^2.
\end{eqnarray}
An Arrott plot is composed of a set of $M^2\,vs.\,H/M$ curves for temperatures near the ferromagnetic transition temperature $T_C$.~\cite{Blundell01,Craig67,yeung} The Curie temperature $T_C$ can be determined by simply finding out which curve passes through the origin of the Arrott plot. According to Eq.\,\ref{eq_A2}, the $M^2\,vs.\,H/M$ curves should be straight lines in an Arrott plot. Unfortunately, magnetic interactions in many ferromagnetic systems are much more complicated than the mean field theory, and consequently the $M^2\,vs.\,H/M$ curves are often not straight lines. Nevertheless, one can still obtain valuable information from the Arrott plot. It was suggested that the slope of the $M^2\,vs.\,H/M$ curves can be used to infer the order of the phase transition, namely a negative (positive) slope is corresponding to the first (second) order phase transition.~\cite{levy}

In this work,  the magnetization measurements were carried out with a very small temperature interval ($\Delta T$=0.5\,K) near $T_C$ in order to obtain sufficient accuracy in the analysis. At each of the temperatures for the Arrott plot shown in Fig.\,2(a) of the main text, the magnetization data were taken after the magnetic field was swept back to zero and the temperature was raised to 300\,K, so that every trace was guaranteed to have the same initial magnetization. The values of internal magnetic field $H$ have been corrected by considering the demagnetization effect, i.e. $H=H_e-NM$, where $H_e$ is the applied (external) magnetic field, $M$ is the magnetization, and $N$ is the demagnetization factor. Following a method reported in Ref.\,~\cite{tsurkan}, $N$ can be evaluated from the low field ($H_e<$100\,Oe) magnetization data in the ferromagnetic phase.

For a second order phase transition, the critical behavior can be described by a series of asymptotic relations:
\begin{subequations} \label{eq:3}
\begin{align}
&M_s(T)=\mathrm{lim}_{H\rightarrow0}M(T)\propto[\frac{T_C-T}{T_C}]^\beta, T<T_C \\
&\chi_0(T)=\mathrm{lim}_{H\rightarrow0}\frac{M(T)}{H}\propto[\frac{T_C}{T-T_C}]^\gamma, T>T_C\\
&M\propto H^\delta, T=T_C
\end{align}
\end{subequations}
Here $\beta$, $\gamma$ and $\delta$ are the critical exponents, $M_s$ is the spontaneous magnetization, and $\chi_0$ is the zero-field limit of the magnetic susceptibility. Following a method described in Ref.\,~\cite{ghosh}, we obtain $M_s(T)$ and 1/$\chi_0$(T) by polynomial fitting the curves in the Arrott plot and extrapolating them to the vertical and horizontal axes, respectively. Fig.\,10 shows the temperature dependences of $M_s(T)$ and 1/$\chi_0$(T), and the corresponding fits to Eqs.\,(A3a) and (A3b).

The critical equations are only valid in a small temperature interval around $T_C$ and in low magnetic fields. Therefore, we limit our fits to a narrow range of temperatures: $|\epsilon|=|(T-T_c)/T|<$\,0.05. The fit shown in Fig.\,\ref{Fig_critical_expn}(a) yields $\beta$=0.361 and $T_C$=105.5\,K. It also shows that the experimental $M_s$ data starts to deviate from the critical relation [Eq.\,A3(a)] at $T$=98\,K, which can be regarded as the lower bound of the critical region. As shown in Fig.\,\ref{Fig_critical_expn}(b), from the fit 1/$\chi_0$(T) data to Eq.\,(A3b), one obtains $\gamma$=1.372 and $T_C$=105.2\,K. Similarly, the upper bound of the critical region is determined to be $T$=117\,K. The In($M$)\,$vs$.\,In($H$) plots at $T$=105.5\,K and 105\,K are shown in Fig.\,\ref{Fig_critical_expn}(c), from which $\delta$=4.83 and 4.85 can be extracted, respectively. Based on this, we obtain $\delta$=4.84 for the average value of the Curie temperature ($T_C$=105.35\,K).

The validity of the above fits can be further confirmed by a method given in Ref.~\cite{Blundell01}. The $M_s(T)$ and 1/$\chi_0$(T) data can be fitted to the following equations:
\begin{subequations} \label{eq:4}
\begin{align}
&Y=\frac{M_s(T)}{dMs(T)/dT}=\frac{T-T_C}{\beta}.\\
&X=\frac{\chi_0^{-1}(T)}{d\chi_0^{-1}(T)/dT}=\frac{T-T_C}{\gamma}.
\end{align}
\end{subequations}

\begin{figure}
\includegraphics[width=8 cm]{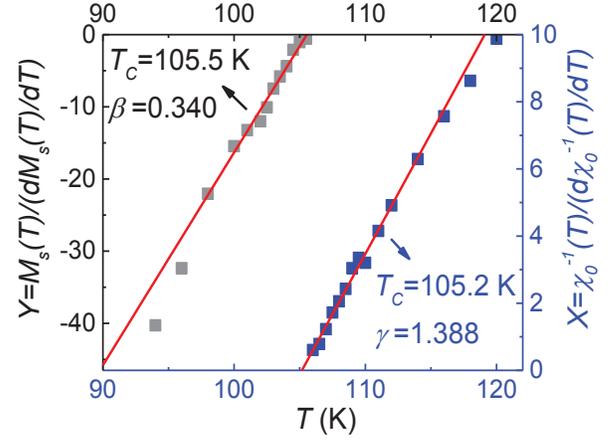}
\caption{(\label{Fig_M_chi}color online). Temperature dependence of the quantities $X$, $Y$ (blue and gray square symbols), defined in the text, and the corresponding linear fits (red lines) to Eqs.\,(A4a) and (A4b). The temperature range with the data points falling onto the linear fits can be regarded as the critical region ($T$=98-117\,K).}
\end{figure}

As shown in Fig.\,\ref{Fig_M_chi}, the obtained values ($\beta$=0.34, $T_C$=105.5\,K and $\gamma$=1.388, $T_C$=105.2\,K) are consistent with the fits with Eqs.\,(4a) and (4b). From this plot, the critical region can be determined to be 98-117\,K, based on where the deviation from the linear relationship takes place. This result is consistent with the analysis shown in Fig.\,\ref{Fig_critical_expn}. The obtained critical exponents ($\beta$=0.361, $\gamma$=1.373, and $\delta$=4.84) are very close to theoretical values of a 3D Heisenberg ferromagnet ($\beta$=0.367, $\gamma$=1.388, and $\delta$=4.78).~\cite{Yanagihara02} This is not surprising since the magnetism near $T_C$ is dominated by the superexchange interactions between Cr$^{3+}$ ions. Similar critical exponents have also been obtained for CdCr$_2$Se$_4$,~\cite{zhang} a compound closer to a magnetic insulator than HgCr$_2$Se$_4$.

\section{ADDITIONAL MAGNETIZATION DATA}

\subsection{Temperature dependence of inverse susceptibility}

\begin{figure}[t]
\includegraphics[width=8.5 cm]{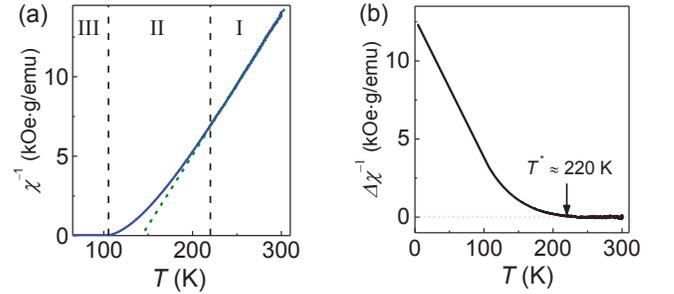}
\caption{\label{Fig_susceptibility}(color online). (a) Temperature dependence of the inverse magnetic susceptibility 1/$\chi$ (solid line) of \emph{n}-HgCr$_2$Se$_4$ and the linear fit in the high temperature range (dashed line). (b) The temperature $T^{\ast}$, at which the deviation from the Curie-Weiss law begins, can be seen more clearly after the linear part of the inverse susceptibility curve is removed, i.e.\ $\Delta\chi^{-1}(T)$=$\chi^{-1}(T)-c(T-T_{\theta})$. The temperature coefficient $c$ and the Curie-Weiss temperature $T_{\theta}$ can be extracted from the linear fit shown in panel (a).}
\end{figure}

In Fig.\,\ref{Fig_susceptibility}(a) we replot the temperature dependence of the inverse magnetic susceptibility 1/$\chi$ of \emph{n}-HgCr$_2$Se$_4$ shown in Fig.\,4(b). The three temperature zones (I-III) are divided by two characteristic temperatures, the Curie temperature $T_C$ and $T^{\ast}$, at which the deviation of magnetic susceptibility $\chi(T)$ begins. In order to obtain $T^{\ast}$ more accurately, the temperature dependence of $\Delta\chi^{-1}(T)$=$\chi^{-1}(T)-c(T-T_{\theta})$, is plotted in Fig.\,\ref{Fig_susceptibility}(b), where $c$ and $T_{\theta}$ are temperature coefficient and the Curie-Weiss temperature, respectively. It clearly shows $T^{\ast}\approx$220\,K$\approx$2.1$T_C$.

\subsection{Carrier density dependences of the Curie temperature}

\begin{figure}
\includegraphics[width=7 cm]{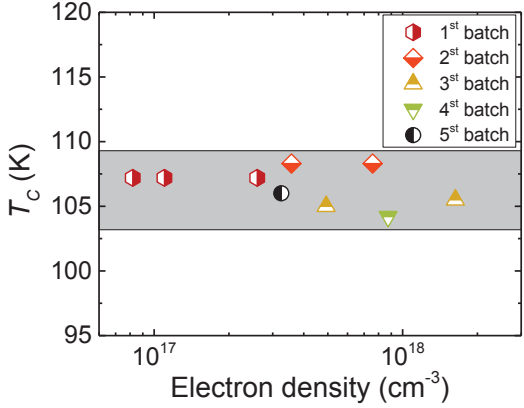}
\caption{\label{Fig_CurieTemp}(color online). Carrier density dependence of the Curie temperature $T_C$ for several batches \emph{n}-type HgCr$_2$Se$_4$ single crystals. The electron densities are extracted from the Hall measurements at 4.2\,K. The Curie temperature $T_C$ is taken to be the kink point of the $M(T,H)\,vs.\,T$ curve with a small magnetic field (typically $H$=100\,Oe) applied. This method gives reasonably accurate $T_C$ values in comparison to those deduced from the critical exponent analysis.}
\end{figure}

 In this work, electron densities in \emph{n}-type HgCr$_2$Se$_4$ samples are in a range of $10^{16}$-$10^{18}$\,cm$^{-3}$, and hence the carrier mediated exchange interaction is not expected to play a significant role due to the strong Cr$^{3+}$-Se$^{2-}$-Cr$^{3+}$ superexchange interactions. This is confirmed by the measurements of the Curie temperatures of the samples from five growth batches. As shown in Fig.\,\ref{Fig_CurieTemp}, there is no obvious dependence of $T_C$ on the carrier density, which is determined by the Hall measurements at liquid helium temperatures.

\section{ADDITIONAL TRANSPORT DATA}

\begin{figure}[b]
\includegraphics[width=8.5 cm]{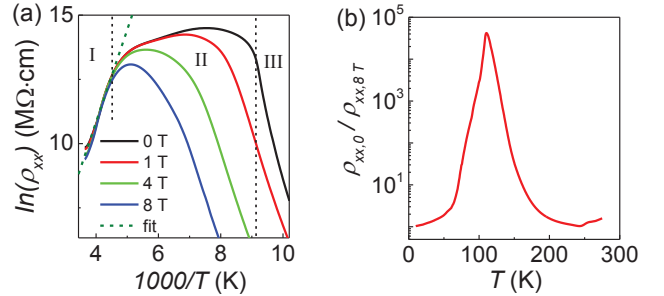}
\caption{\label{Fig_CMR2}(color online). Transport properties of an \emph{n}-type HgCr$_2$Se$_4$ single crystal (Sample\,B). (a) Temperature dependences of the longitudinal resistivity plotted as ln$(\rho_\mathrm{xx})\,vs.\,1/T$ for magnetic fields $\mu_0H$=0,\,1,\,4,\,8\,T. The linear fit (dashed line) for $T>$220\,K yields a thermal activation gap $\Delta$=0.28\,eV. The transport characteristics in temperature zones (I) $T>T^*$, (II) $T_C<T<T^*$ and (III) $T<T_C$ resemble those of sample\,A [See Fig.\,4(a) in the main text]. (b) $T$-dependence of the MR ratio $\rho_\mathrm{xx,0}/\rho_\mathrm{xx}(H)$ with $\mu_0H$=8\,T. The maximum value is 4.2$\times10^4$, which is located at $T$=110\,K.}
\end{figure}

Electron transport measurements have been carried out on several batches of \emph{n}-type HgCr$_2$Se$_4$ single crystals. As long as the carrier density is on the order of $10^{18}$\,cm$^{-3}$, transport properties similar to those of Sample A (presented in the main text) can be observed. In Fig.\,\ref{Fig_CMR2}, we plot the transport data from Sample B. In addition, we also observed the CMR effects with MR ratios up to several orders of magnitude in samples with carrier densities on the order of $10^{17}$\,cm$^{-3}$ or lower. In these low-carrier-density samples, however, the resistivities near the magnetic phase transition are too high to perform reliable four-point measurements. We therefore restricted our focus on samples with electron densities on the order of $10^{18}$\,cm$^{-3}$ in the present work.


\section{TEMPERATURE DEPENDENCE OF LATTICE CONSTANT}

\begin{figure}
\includegraphics[width=7 cm]{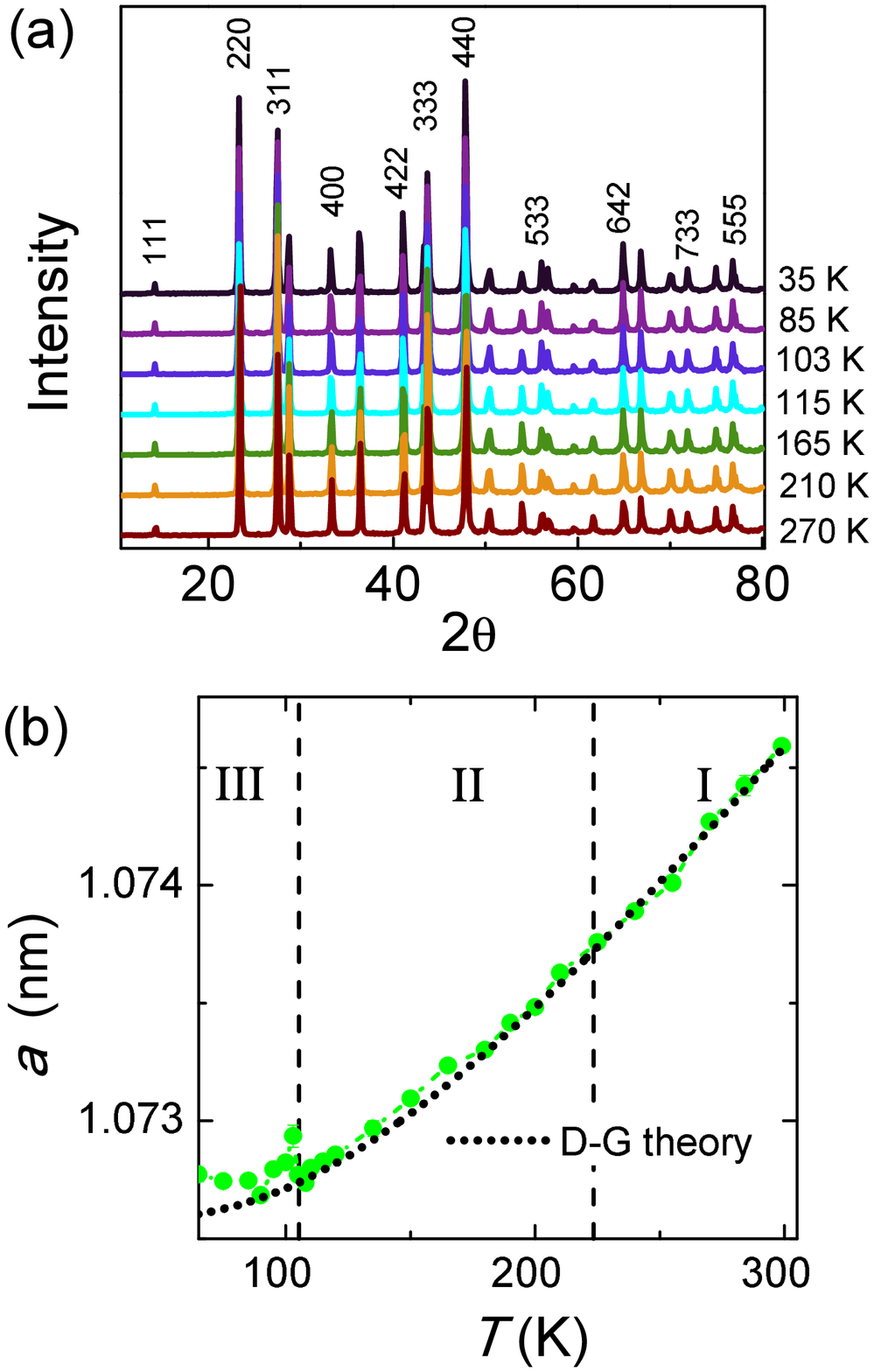}
\caption{\label{Fig_lattcie_const}(color online). (a) X-ray diffraction (XRD) spectra of HgCr$_2$Se$_4$ in a temperature range of 35-299\,K. (b) Temperature dependence of the lattice constant extracted from the XRD measurements (solid circles), which is compared with the Debye-Gruneisen (D-G) theory (dotted line).}
\end{figure}

The crystalline structure of HgCr$_2$Se$_4$ was determined by using a variable temperature X-ray diffractometer with a temperature range of 35-299\,K. The X-ray diffraction spectra are shown in Fig.\,\ref{Fig_lattcie_const}(a) of a powder sample which was prepared by grinding many HgCr$_2$Se$_4$ single crystals to warrant sufficient accuracy. The temperature dependence of the lattice constant that deduced from analysis of the X-ray diffraction spectra are shown in Fig.\,\ref{Fig_lattcie_const}(b). It can be well described by the Debye-Gruneisen theory except a slightly deviation around $T_C$.  This indicates that the magnetic polarons in \emph{n}-HgCr$_2$Se$_4$ are also different from the small polarons observed in perovskite manganites. The Jahn-Teller distortion in manganites produces stable, spatially confined dielectric polarons.~\cite{Salamon01,Teresa97} The hopping transport related to these small polarons can, however, survive at very high temperatures (e.g.\ 1000\,K),~\cite{Worledge96} at which the spin correlation effect is negligible.~\cite{Salamon01} In \emph{n}-HgCr$_2$Se$_4$ the trivalent Cr ions are not susceptible to the Jahn-Teller distortion, and the small dielectric polaron scenario is thus unlikely to be relevant to the results obtained in this work.


\end{appendix}

\end{document}